\documentclass[epsfig,pra]{revtex4}
\usepackage{graphicx}
\usepackage{amssymb}
\usepackage{amsmath}
\usepackage{graphicx}
\usepackage{epstopdf}
\usepackage{subfigure}
\newcommand{\eq}{\begin{eqnarray}}
\newcommand{\en}{\end{eqnarray}}

\newcommand{\be}{\begin{equation}}
\newcommand{\ee}{\end{equation}}
\newcommand{\bea}{\begin{eqnarray}}
\newcommand{\eea}{\end{eqnarray}}

\newcommand{\ket}[1]{\ensuremath{| #1 \rangle}}

\begin{document}

\title{Natural Light Harvesting Systems: Unraveling the quantum puzzles}


\author{A. Thilagam}
\email{thilaphys@gmail.com}
\affiliation{Information Technology, Engineering and Environment,\\ 
University of South Australia, Australia
 5095.}
\date{\today}


\begin{abstract}
In natural light harvesting systems, the sequential quantum events of photon absorption by specialized
biological antenna complexes, charge separation, exciton formation and energy transfer to  localized reaction centers
culminates in the conversion of solar to chemical energy. 
A notable feature in these processes is the exceptionally high efficiencies ($>$ 95\%) at which
excitation is transferred from the illuminated protein complex site  to the 
 reaction centers.  The  high  speeds of excitation propagation within a   system of interwoven biomolecular network structures, is
 yet to be replicated in artificial light harvesting complexes.  A clue to unraveling the quantum puzzles of nature may lie in the observations of long lived coherences lasting several picoseconds in the electronic spectra of photosynthetic complexes which occurs  even in noisy environmental baths. However the exact nature of the association between the high energy propagation rates and strength of quantum coherences remains largely unsolved. 
A  number of experimental  and theoretical studies have been devoted
to unlocking the links between quantum processes and information protocols, 
in the hope of finding the answers to nature's puzzling mode of energy propagation. 
This review presents recent developments in quantum theories, and 
links  information-theoretic aspects with  photosynthetic light-harvesting
processes in biomolecular systems. There is examination of various attempts to pinpoint the processes that
 underpin coherence features arising from the light harvesting activities of biomolecular systems, with particular emphasis on the effects 
that factors such non-Markovianity, zeno mechanisms, teleportation,  quantum predictability
 and the role of multipartite states  have on the quantum dynamics of biomolecular systems. 
A discussion of  how quantum thermodynamical principles and agent-based modeling and simulation approaches can 
 improve our understanding of natural photosynthetic systems is included.
\end{abstract}
\maketitle

\section{Introduction}
Photosynthesis is an important quantum process which 
contributes significantly to the world's yearly biomass yield.
During  photosynthesis,  light absorption followed by charge separation
 and  efficient energy transfer to a reaction center (RC) are performed by specialized  pigment-protein
(LH) complexes \cite{knox,p1,p1a,p3,may,ame,cogdell}. 
The  conversion of solar energy to chemical energy takes place
at the reaction center, with the chemical reactions providing the driving force
for the adenosine triphosphate
complex,  ATP. This co-enzyme forms the basis for critical cellular processes
needed for survival of the supported species.
The theory of excitonic energy  transfer in  light harvesting systems 
(LHS)  has been a topic of  interest  over several decades 
  \cite{knox,p1,p1a,p3,may,p0,fors,macro,lloyd,thilchem2,silbey,ghosh,olbri,pala1,pala2,pala,rit,hoyer}.
The light harvesting complexes from different species vary in their
 structural arrangements, but possess the common attribute
of enabling excitation propagation even in  adverse, noisy environments.

\begin{figure}[htp]
\centering
\subfigure{\label{f1}\includegraphics[width=5.5cm]{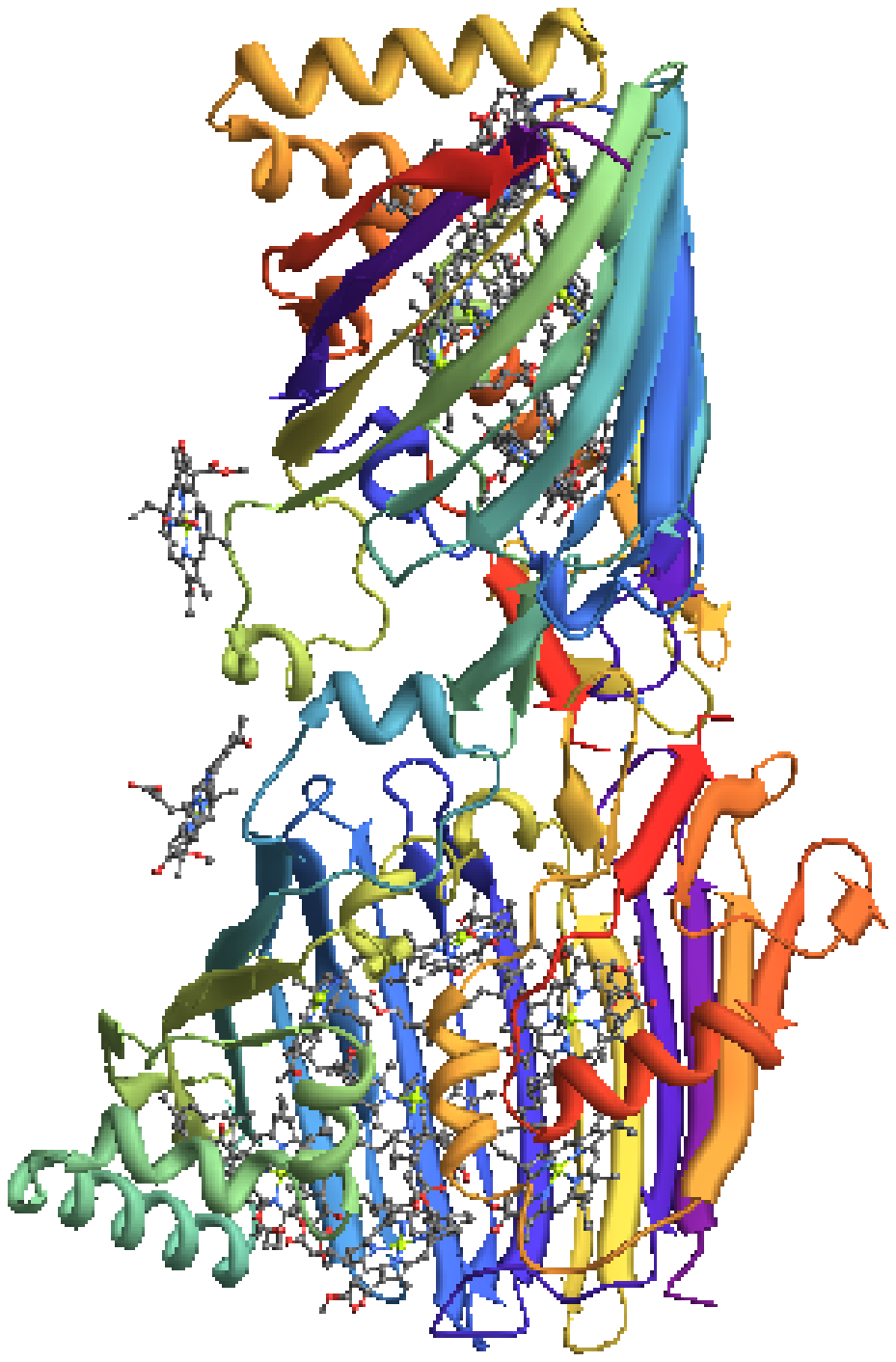}}\vspace{-1.1mm} \hspace{1.1mm}
\subfigure{\label{f2}\includegraphics[width=4.20cm]{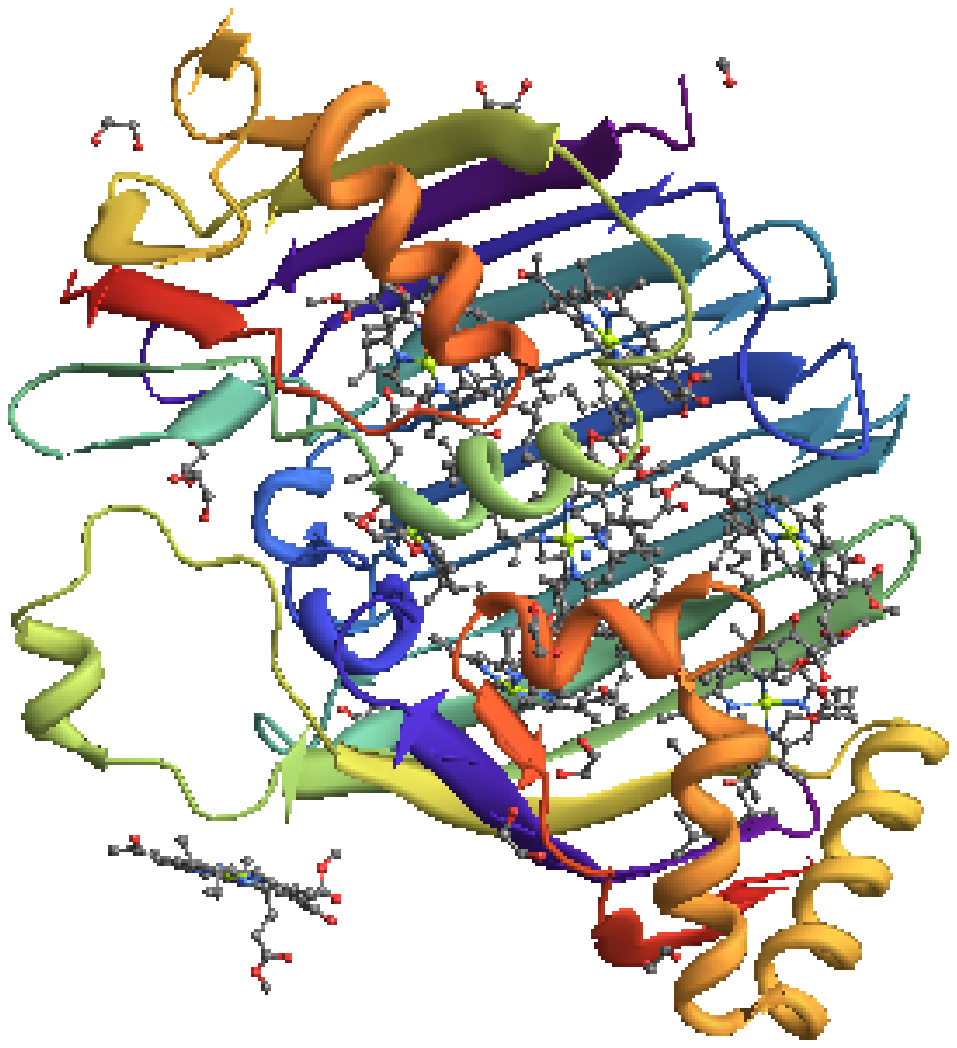}}\vspace{-1.1mm} \hspace{1.1mm}
\caption{(Left) Crystal structure of the Fenna-Matthews-Olson protein complex from
the  sulphur bacteria  species, \textit{Chlorobaculum Tepidum}
obtained using the pdb code:3ENI
from http://www.rcsb.org/pdb/download \cite{tron,tron1}.
\quad
(Right) Crystal structure of the Fenna-Matthews-Olson protein complex from
the sulphur  bacteria  species, \textit{Prosthecochloris Aestuarii}
obtained using the pdb code:3EOJ.}
 \label{lhs}
\end{figure}

\begin{figure}
\centering
\includegraphics[width=8cm]{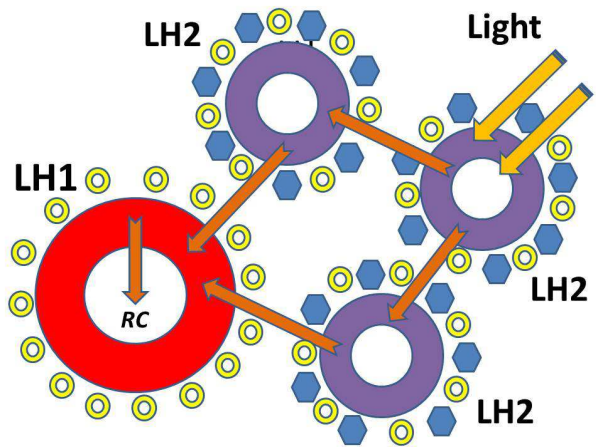}
\caption{Schematics  of the photosynthetic apparatus of purple bacteria,
in which the yellow carotenoids and different BChls protein complexes (blue, purple and red) 
 absorb solar energy. The excitation is   transferred via excitonic propagation (indicated by orange arrows)
from the LH2 antenna complexes to the LH1 complex, and then to 
 the reaction center (RC) where energy conversion to chemical form occurs.
 This figure is based on the photosynthetic apparatus scheme which appear
in Ref.\cite{rit}}
 \label{pur}
\end{figure}

The absorption of a photon results in the formation of
an excited  quasi-particle  known as the exciton (correlated electron-hole pair), on one or more
optically active molecular sites occupied by molecular complexes, known as chromophores. 
One well studied pigment-protein complex
is the Fenna$-$Matthews$-$Olson (FMO) complex  
of the  green sulfur bacteria \cite{tron,tron1,flem2,fenn,cam,engel,reng,wend,mark} shown in 
Fig.~\ref{lhs}  for two species of the bacteria. The FMO
acts as the prototypical system for quantum studies during  photosynthetic activities,  mainly due to 
numerical tractability of its known crystallographic structure, and availability of a
 wide range of experimental results for this system.
The FMO complex trimer is made up of  three symmetry equivalent
 monomer subunits, with each subunit constituting eight bacteriochlorophyll
(BChl)a molecules (Fig.~\ref{fmo}) supported by a cage  of protein molecules . 
The chromophore sites numbered $3$ and $4$ are located  near the reaction center,
and thus are closely linked to the sink region where energy is released,
while chromophore sites $1$ and $2$  are strongly coupled to each other dissipating energy via  site $3$.
The sites $1$, $6$, and $8$ are located at the baseplate which connects
to the chlorosomes  that receive electronic excitations. The eighth chromophore
 plays a critical role in the topological connectivity of
large molecular structures, 
and in this regard is critical to the existence  of multipartite
states. The  Fenna-Matthews-Olson (FMO) protein  complex
is sandwiched  between the large peripheral chlorosome antenna complex and the reaction center (RC), 
and excitation originating at the antenna site
propagates through the FMO to the reaction center. 
In the case of the  purple photosynthetic
bacteria, there exist two different types of light harvesting molecular system
units with ring-like structures,  known as LH1 and LH2 (Fig.~\ref{pur}).
The core LH1 complex is directly linked to the RC by surrounding it, 
while the peripheral LH2 complexes play an important role by
 transferring  energy to the LH1 complex. Hence  the photosynthetic apparatus of both the green sulphur and
purple bacteria are formed from molecular complexes with two  distinct roles based
on whether the molecular complex is directly or indirectly  linked to
the reaction center. 

 The FMO complex exhibits strong quantum beating lasting ($>$ 600 fs),
which currently is not fully accounted for by any theoretical  predictions. 
 The long-lasting coherent dynamics are indicated as cross-peak oscillations in two-dimensional
echo-spectra results \cite{flem2,cam,engel,reng}. 
The  exceptionally high efficiencies ( $>$ 95\%) at which
 excitation  propagates between the  
light harvesting complexes before reaching the  reaction center (RC) 
pigment-protein complex \cite{p1,p3,may} is yet 
to be realized in artificial light harvesting complexes.
Two-dimensional photon-echo  based 
experiments \cite{engel} in which 
the time delays of fast pulses are manipulated to provide  a 
map of  excitation and emission frequencies at select durations of  time delay, is a 
 reliable   technique for examining
energy transfer processes. Recent progress
in photon based instrumentation involving detection and counting techniques,
has facilitated the reconstruction of a system's  state via the
 promising  quantum-state tomography approach, which
allow quantum states and their associated density matrices to be estimated to
a high degree of accuracy \cite{niel,tomog,tomoprl,excitomo,wein,walm}.

\begin{figure}
\centering
\subfigure{\label{f1}\includegraphics[width=5.55cm]{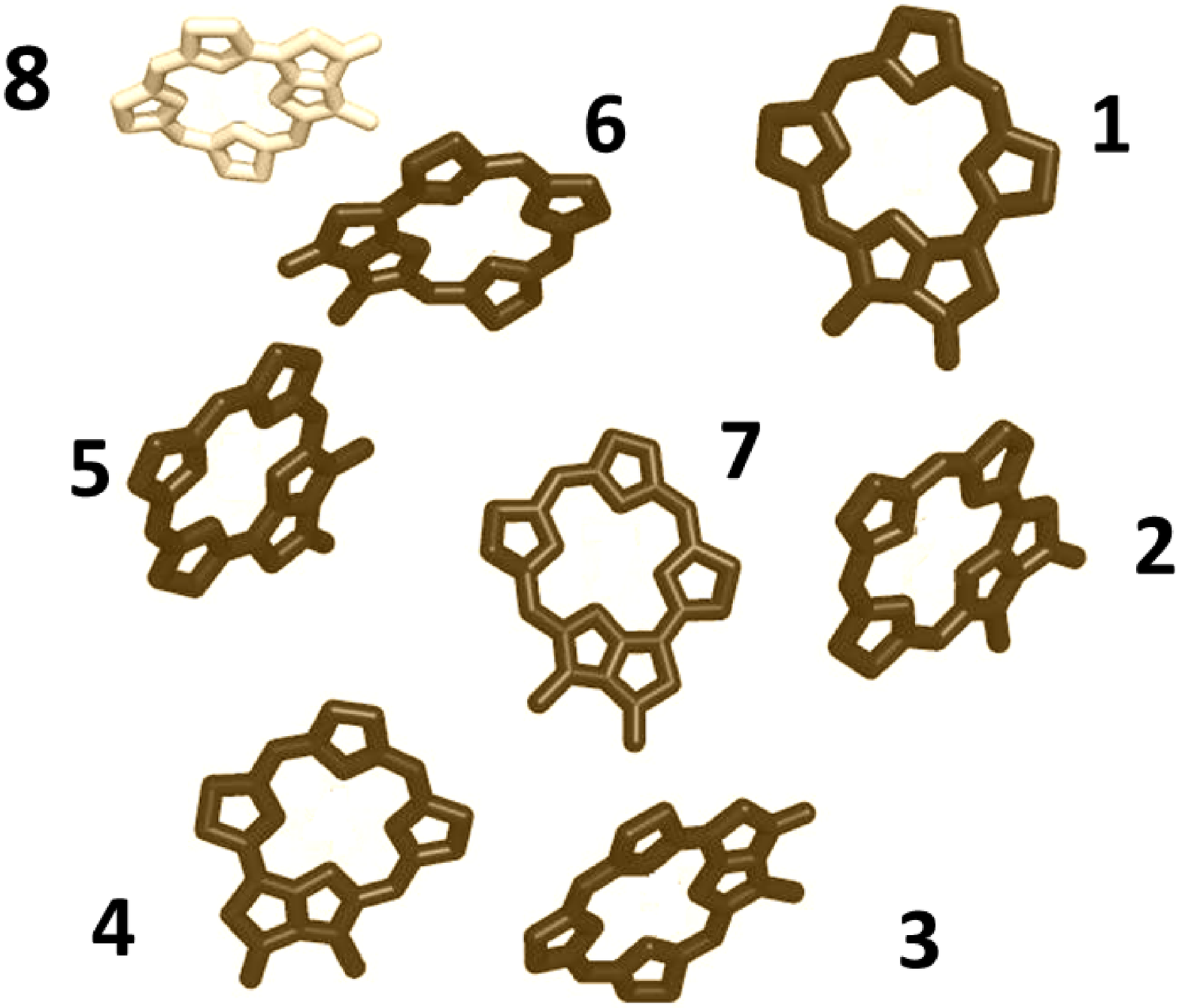}}\vspace{-1.1mm} \hspace{1.1mm}
\subfigure{\label{f2}\includegraphics[width=5.60cm]{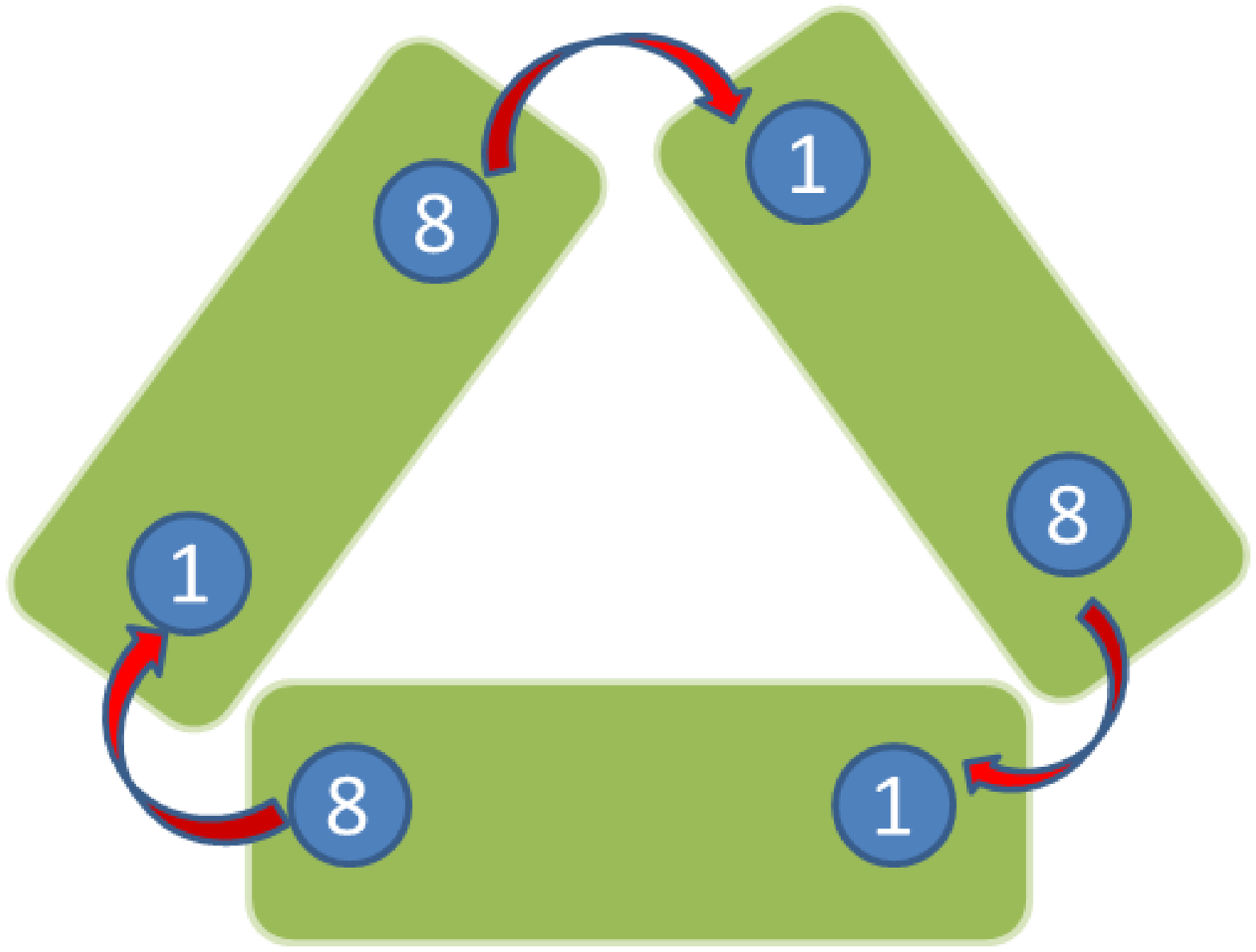}}\vspace{-1.1mm} \hspace{1.1mm}
\caption{(left) FMO complex  monomer subunit constituting eight bacteriochlorophyll
(BChl)a molecules and (right) Simplified trimer configuration,
 in which the eighth BChl-a is positioned close to the first
chromophore molecule of a neighbouring monomer. This results
in strong inter-monomer interactions for  excitations present at sites 8 and 1.}
 \label{fmo}
\end{figure}

\section{Exciton dynamics of  the FMO complex}\label{fmo}

In this Section, we examine the exciton  dynamics of the 
 Fenna-Matthews-Olson (FMO) complex using  
the site energies  of the   BChl molecules based on experimental findings of
 its  crystal structure \cite{madjet}. The
coupling energies between BChl molecules derived using the dipole-dipole approximation
 are obtained from Ref.\cite{moix}. Photon  absorption by a BChl molecule causes a 
newly created exciton to spread rapidly to the adjacent units while interacting 
with a bath of phonons. The  propagating exciton  may be 
distinguished  from  an excitation localized at a specific BChl site.
 Excitonic states can be represented as qubit states 
 by  associating a qubit with  the presence (or absence) of an exciton at a
specific energy level determined by the exciton Hamiltonian. Each energy level is then modeled as a two-level system,
using $\ket{0}_{e_i}$ ($\ket{1}_{e_i}$) which denotes
the absence (presence) of an excitation at the specified energy level $j$. 
 One or more  BChl sites may contributes predominantly to a specific  qubit state,
as will be shown shortly. 
 We  employ the  effective exciton Hamiltonian defined in the basis of the Q$_y$
one-exciton states,  where the   Q$_y$ bandwidth is associated with the lowest 
excited state of the BChl molecule.  Using known site and
coupling energies between    BChl molecules \cite{madjet,moix},
the Hermitian Hamiltonian, $\hat H_{ex}$  of the FMO complex with eight BChl sites is given  in  units of cm$^{-1}$
as

\begin{equation}\label{FMOcom} \hat H_{ex}= \left(
\begin{array}{cccccccc}
 310.0 & -97.9 & 5.5 & -5.8 & 6.7 & -12.1 & -10.3 & 37.5 \\
 -97.9 & 230.0 & 30.1 & 7.3 & 2.0 & 11.5 & 4.8 & 7.9 \\
 5.5 & 30.1 & 0.0 & -58.8 & -1.5 & -9.6 & 4.7 & 1.5 \\
 -5.8 & 7.3 & -58.8 & 180. & -64.9 & -17.4 & -64.4 & -1.7 \\
 6.7 & 2.0 & -1.5 & -64.9 & 405. & 89.0 & -6.4 & 4.5 \\
 -12.1 & 11.5 & -9.6 & -17.4 & 89. & 320. & 31.7 & -9.7 \\
 -10.3 & 4.8 & 4.7 & -64.4 & -6.4 & 31.7 & 270 & -11.4 \\
 37.5 & 7.9 & 1.5 & -1.7 & 4.5 & -9.7 & -11.4 & 505.0 \\
\end{array}
\right). \end{equation}

\begin{figure}
\centering
\includegraphics[width=8cm]{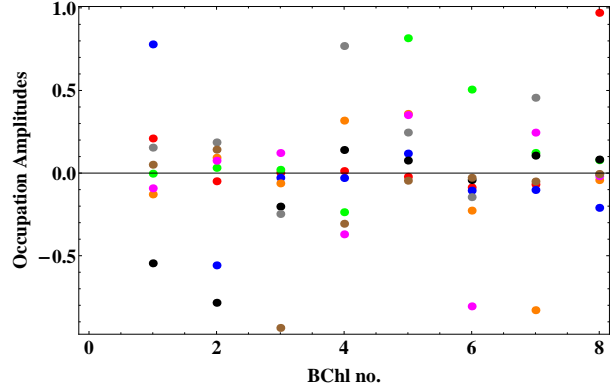}
\caption{The probability amplitudes
of exciton occupation at the  eight BChls of the FMO complex, evaluated using the
Hermitian Hamiltonian, $\hat H_{ex}$ in Eq.\ref{FMOcom}. The eight exciton qubits are  coded using
different colored dots. }
 \label{simult}
\end{figure}

The eigen-energies and   eigenstates of excitations described by $\hat H_{ex}$ (\ref{FMOcom}) can be evaluated  
to reveal the eight excitonic qubits 
(labelled in order of increasing energy) as shown in Table  \ref{tblqu}.
Each exciton state appears to be  present at two or more   BChl sites.
For instance, the lowest energy exciton qubit state $\ket{1}_{e_1}$ with energy $-24.2$ cm$^{-1}$ 
 is localized at BChls 3 and 4, while the qubit state $\ket{1}_{e_2}$ with energy $143.9$ cm$^{-1}$ 
 is localized predominantly at BChls 4 and 7. The qubit state $\ket{1}_{e_8}$ with the maximum energy $514.3$ cm$^{-1}$ 
 has maximal localization at BChls 8. In Fig \ref{simult}, the probability amplitudes
of exciton occupation at the different BChl site is shown, highlighting
that each exciton qubit state 
has simultaneous occupation at  several BChl sites,  an intrinsically quantum attribute.
Fig \ref{simult} shows that at some BChl sites, there appears to be clustering of qubits (1, 3, 8)  with one
qubit appearing with a comparatively large probability amplitude. Interestingly, BChl 8 shows
non-trivial amplitudes for some qubits, highlighting its important role during photosynthesis.
Earlier works have generally focussed on a model  of FMO complex with seven BChl sites,
generally ignoring the eight BChl site.
Due to the  superposed
state involving all the eight BChls of the FMO complex,  each exciton
can be expressed as a   linear combination of the excited state wave functions of
the individual  BChl.   We discuss further the implications of the unique state of existence of 
quantum superposed states  under the  ``coherent propagation scheme" discussed
in Section \ref{co}.

\begin{table}
\begin{center}
\caption{
\label{tblqu}
Delocalized exciton qubit states provided
as linear combinations of   probability occupation amplitudes associated with
the eight BChl sites. The excitonic qubit states are labeled according to decreasing energies (cm$^{-1}$) for the
FMO complex.
\newline
}

    \begin{tabular}{| r | r | r | r | r  | r |  r |  r | r | r |r|}
    \hline
     &$E_x$(cm$^{-1}$) &  BChl 1&  BChl 2&  BChl 3&  BChl 4&  BChl 5& BChl 6&  BChl 7& BChl 8  \\ \hline 
 ~ $\ket{1}_{e_8}$  &514.3 \quad \quad &0.209 & -0.049 &  0.002 &  0.012 &  -0.022 &  -0.087 &
 -0.069 &  0.969\\
 ~ $\ket{1}_{e_7}$ & 478.4  \quad \quad &
 -0.004 &  0.032 &  0.020 &  -0.237 &  0.816 & 0.505 &  0.122 &  0.078 \\ 
 ~ $\ket{1}_{e_6}$  & 374.1  \quad \quad & 0.779 &  -0.558 &  -0.029 &  -0.030 &  0.118 &  -0.106 &  -0.102 &
 -0.210 \\
~ $\ket{1}_{e_5}$
 & 303.7  \quad \quad & -0.129 & 0.093 & -0.062 & 0.318 & 0.358 & -0.227 & -0.830 &
 -0.042  \\ 
 ~ $\ket{1}_{e_4}$  & 262.4  \quad \quad & -0.092 &  0.075 &  0.121&  -0.371 &  0.351 &  -0.806 &
 0.245 &  -0.019 \\ 
 ~ $\ket{1}_{e_3}$  & 167.3  \quad \quad & -0.546 &  -0.784 &  -0.203 &  0.140 &  0.076 &  -0.042 &
 0.106 &  0.082 \\ 
 ~ $\ket{1}_{e_2}$   & 143.9  \quad \quad & 0.154 &  0.186 &  -0.247 &  0.769 &  0.245 &  -0.146 &
 0.456 &  -0.008  \\ 
 ~ $\ket{1}_{e_1}$  &-24.2  \quad \quad &  0.051 &  0.142 &  -0.937 &  -0.307 &  -0.046 &
 -0.028 &  -0.051 &  -0.005 \\ \hline
    \end{tabular}
\end{center}
 \end{table}

The system of the two-level excitonic qubit  in contact with a bath of phonons is described by the Hamiltonian
\begin{equation}
\label{ham}
{H}=\omega_0{\sigma}_+{\sigma}_- \sum_{\bf q} \hbar \omega_{\bf q} \,
b_{\bf q}^{\dagger}\,b_{\bf q} + \sum_{\bf q}   \lambda_{_{\bf q}} \,
 \left ( {\sigma}_- \,b_{\bf q}^\dagger +{\sigma}_+ \,b_{\bf q} \right ),
\end{equation}
where ${\sigma}_+=|1\rangle\langle0|$ and
${\sigma}_-=|0\rangle\langle1|$, 
are the respective Pauli raising and
lowering operators of the exciton with   transition frequency, $\omega_0$. 
In order to introduce some degree of tractability, 
each qubit is considered as coupled to its own reservoir of phonons which appear as 
the second term on the right hand side of  Eq. (\ref{ham}). The terms,
$b_{\bf q}^{\dagger}\,$ and $b_{\bf q}\,$
are the respective  phonon creation and  annihilation 
operators with wave vector ${\bf q}$. The last term of Eq. (\ref{ham}) denotes  the qubit-oscillator interaction Hamiltonian which we assume to be linear combination of the phonon operators. 
 $\lambda_{_{\bf q}}$ is the coupling between the qubit and the environment 
and is characterized by  the spectral density function, such as,
$J(\omega)=\sum_{\bf q}\lambda_{_{\bf q}}^2\delta(\omega-\omega_{\bf q})$.

Analytical expressions for the dynamical evolutions of the excitonic qubit can be obtained by considering Lorentzian forms for  the spectral density arising from the interaction between the  BChl exciton and  the phonon bath 
  \be 
\label{spectral}
J(\omega)=  \frac{1}{2\pi }\gamma_0 \; \left( \frac{\Delta \omega}{2} \right)^2
 \frac{1}{(\omega_0  - \delta -\omega)^2+\left( \frac{\Delta \omega}{2} \right)^2}
\ee
whose  central peak is detuned from the exciton transition frequency $\omega_0$ 
 by an amount $\delta$. 
The full width at half-maximum 
$\Delta \omega$ is linked to the  reservoir correlation time $\tau_B$ via $\Delta \omega = \frac{2}{ \tau_B}$.
The parameter $\gamma_0$ is associated with 
 the relaxation time scale $\tau_R$ of the exciton via  the relation $\tau_R=\gamma_0^{-1}$. 
The reservoir correlation and exciton relaxation parameters, which are
dependent on the temperature, may be 
employed as semi-empirical variables to reduce the computational times
needed to examine the dynamics of excitons.

The excitonic  qubit decays  to 
oscillator states in the reservoir undergoing
a transition from  the excited state
$\ket{1}_{e}$ to the ground state $\ket{0}_e$.  The initial state of the qubit with  its corresponding 
reservoir in the vacuum state appear as
\begin{equation}
|\phi _{i}\rangle =|1\rangle _e\otimes
\prod_{k=1}^{N'}|0_{k}\rangle _{\mathrm{r}}=
|1\rangle _e\otimes
\ket{{ 0}}_{\mathrm{r}}, 
 \label{initial}
\end{equation}
where $\ket{{ 0}}_{\mathrm{r}}$ implies that all $N'$ wavevector modes 
of the reservoir are unoccupied in the initial state.
The subscripts $e$ and $r$ refer to the excitonic qubit and the corresponding reservoir
respectively.  $|\phi _{i}\rangle$    undergoes subsequent decay of the 
following form
\begin{equation}
|\phi _{i}\rangle \longrightarrow
 u(t) \; \ket{1}_e
\ket{{ 0}}_{\mathrm{r}} + v(t) \; \ket{0}_e
\ket{{ 1}}_{\mathrm{r}},  
\label{fstate}
\end{equation}
We consider that $\ket{{ 1}}_{\mathrm{r}}$ denotes 
a  collective state of  the reservoir
\begin{equation}
\label{crstate}
|{ 1}\rangle _{\mathrm{r}}=\frac{1}{v(t)}
\sum_{n} \lambda _{\{n\}}(t)|\{n\}\rangle ,
\end{equation}
where $\{n\}$ denotes an occupation scheme in which
$n_i$ oscillators with wavevector $k=i$ are present in the reservoir.
We define the state $|\{n\}\rangle$ as 
$|\{n\}\rangle =|n_0,n_1,n_2...n_i..n_{N'}\rangle$.
In the  collective reservoir state,  the phonon oscillators
can be  present at all allowed modes, including
simultaneous excitation of several  phonon states. 

The time-convolutionless (TCL) projection operator 
technique \cite{breu} is employed to obtain the coefficient $u(t)$ (see  Eq. (\ref{fstate})),
\begin{equation}
\label{laplace}
\dot{u}(t)=-\int\limits_0^t \; \int {d\omega J(\omega )} \exp
[i(\omega_0-\omega)(t - t_1 )] \; u(t_1)dt_1,
\end{equation}
The  Laplace transformation of Eq. (\ref{laplace})
 with initial condition $u(0)=1$ and the form of spectral density in Eq. (\ref{spectral})  yields the solution \cite{breu}
  \begin{equation}
\label{ana}
u(t) = e^{-(\Delta \omega/2- i\delta)t/2)}\left [ \cosh \left (\frac{\xi t}{2} \right) 
+ \frac{\Delta \omega/ 2 - i\delta }{\xi }\sinh \left ( \frac{\xi t}{2} \right )\right],
\end{equation}
where $\xi=\sqrt{(\Delta \omega/2-i\delta)^2-\gamma_0 \;\Delta \omega}$.
The exciton population difference, $\Delta P$, a signature of coherence, has been evaluated \cite{thilchem2} 
using  $|u(t)|^2$ (Eq. (\ref{ana}). The results in Ref.\cite{thilchem2} show that
at small  $\Delta \omega/2 \sim$20 cm$^{-1}$  or large
reservoir correlation times and large $\gamma_0$ (or small exciton relaxation times), 
there is increased time period (up to 1 ps)    over which the population difference, $\Delta P$
remains resilient. This appears to be in agreement experimental results
of  the  FMO complex of  \textit{P. aestuarii}  \cite{lorenExpt}. We have presented a model of exciton dynamics based on a convenient (and semi-empirical)  Lorentzian form of the spectral density (Eq.\ref{spectral}) and employed the collective state of  the reservoir given in Eq.\ref{crstate}.
While we have focussed on the  quantum dynamics of the FMO complex,
the prototypical features of this system
related to energy propagation provides a reference point from which
the quantum properties of  other energy harvesting molecular systems may be examined.
In the next Section,
we discuss the  challenges involved in the modeling
 excitation  propagation in organic molecular systems,
with  emphasis on  the unique features presented by  the coherent propagation model, which also appears
in the results of Table  \ref{tblqu}.

\section{Challenges in theoretical modeling and the coherent propagation model}\label{co}

A major difficulty  in the theoretical modeling of the energy transfer process
 arises due to  the almost equivalent match in  energy scales ($10-100\ {\rm cm}^{-1}$) of two competing
processes: exciton delocalization and
decoherence due to lattice vibrations in light-harvesting systems.
The Markovian approximation, in which an infinitely short correlation reservoir
time  is assumed, becomes unreliable in the time regime where there is delicate interplay 
between the two main  competing processes. 
The  complexity of modeling large complex molecules against a background of noisy processes
also presents numerical challenges in verifying the
 links between the  coherence times and 
the  excitation energy transfer times. In general simpler forms 
of the spectral density of the bath oscillators are assumed 
(see Eq.\ref{spectral}) to provide tractability, at the expense of accuracy of computed results.
In a recent work \cite{mark},  a signal processing technique was employed 
 to construct  reservoir systems models with increased accuracy and decreased computational cost.
The electronic degrees of freedom of 
molecular complexes involved in photon excitation (at the 
solar illumination site) and those associated with the excitation transfer 
 between chromophores are noted to be intricately mixed with 
environmental degrees of freedom (reservoir, impurities)
 present in the system. 
The validity of the usual procedure of eradicating 
environmental variables to construct a 
pure reduced density matrix is still debatable, with regard to non-Markovian effects
(discussed in Section \ref{mar}). Consequently, many findings
related to photosynthetic systems involve model systems  which are subject to assumptions,
introducing artifacts  which 
influence the final outcome of the results.

In current theoretical studies of photosynthetic systems, the once popular classically derived exciton
 hopping model  \cite{knox,p0,fors} has  been replaced by modern approaches
 based on the quantum coherence properties of the exciton 
\cite{lloyd,alex,silbey,ghosh,sar,hierac,shaun}. 
In the coherent propagation scheme which we  briefly examined in  Section \ref{fmo},  the exciton is 
 modeled as delocalized excitation  which spans the real crystal space \cite{Davy,Craig,toy,suna,thilaold}
as an extended entangled system \cite{thilapra,thilma,thilarxiv}.
The exciton is considered to be 
 in a state of existence at several lattice or BChl sites (Fig \ref{simult}),
traversing multiple paths simultaneously, and undergo  continuous interferences. 
This model is based on  Feynman's ``sum over histories" rule which incorporates all possible
paths between two points, including phase interferences. In essence, it is these interferences
which give rise to the uniquely quantum behavior present in many solids. 
In a molecular crystal, the degree of exciton delocalization is influenced by the environmental
 bath of   phonons and other dissipative factors (impurities and trapping centers). 
In optimal situations, the superposed states may be considered to direct the exciton  to find the most efficient route to 
 the site where energy conversion takes place.
 The system  ``checks"  many states simultaneously, and selects the ``winner" sites,
this idea was proposed in an earlier work which examined the  Grover-like search by excitonic 
states \cite{thilapra}. In the original implementation of the  Grover search \cite{grov1,grov2},
 the location of an item in an unsorted database containing N elements is attempted. While this 
process requires $O(N)$ steps during classical computation,
Grover \cite{grov1} showed that only $O(\sqrt{N})$ steps are required in the presence
of quantum interference effects. In Ref.\cite{thilapra} however, the subtle links 
between the appearance of the excitonic superposed states  and experimental observations  of
quantum coherences  in   noisy  environments  have not been rigorously shown.

The important role of  spin  dynamics \cite{spin1,spin2,spin3} of molecular systems
 during the photosynthetic process needs mention. 
Spintronic systems involve spin-up and spin-down charge carriers, 
and the information encoded in spins may be used to 
 exploit quantum coherence aspects of the global system. When exposed to sunlight, the absorption of a photon may induce intra-molecular and inter-molecular transfer of  electrons and excitons,
giving rise to radical-molecules with unpaired electrons or a different orientation of spin structures. Importantly, these subtle alterations in spin structure of the biomolecules
enable different routes of energy propagation via changes in the chemical environment,
and spin-specific responses to the lattice vibrations \cite{thilohe}. To date, the 
 role of molecular spintronics in light harvesting properties has not been fully examined.
From a quantum information theoretic perspective, photosynthetic system appear to present
a viable platform in which to explore the quantum coherence properties  arising from spin dynamics of an
extended biological network system. 

The search is  ongoing for a comprehensive theory which
can fully account for the observed long-lasting coherent dynamics in natural systems
with light harvesting potential.
There may exist specific subspaces 
of information theoretic entities, which  give rise to the
strong coherences noted in experimental works,
despite the presence of a background of oscillators.
Lately, we  examined  links between the observed coherence
and the environment-assisted transfer mechanisms based on the Zeno-effect \cite{thilzeno}.
The quantum Zeno effect is the retarded time evolution 
of a quantum state subject to frequent measurements, with the reverse
effect resulting in  enhancement of quantum dynamical evolution,  known 
as anti-Zeno effect. The results in Ref.\cite{thilzeno}
indicate that sites of a  dissipative nature act  indirectly as  detectors,
to induce anti-Zeno-like effects, thereby facilitating 
information feedback into the specific partitions in the biomolecular system.
In an earlier  work \cite{fujii}, 
 repeated measurements in disordered systems was seen to   induce a quantum 
anti-Zeno effect with ability to enhance quantum transport  under certain conditions.
These works \cite{thilzeno,fujii} highlight the importance of the Zeno mechanisms in photosynthetic
processes, and the Zeno mechanism-measurement theory point
to the critical role  of quantum mechanical principles during photosynthesis.

\section{Approaches to examining the quantum dynamics of  photosynthesis }{\label{other}}
Several studies of excitation transfer in  quantum systems \cite{red,weiss,breu} have employed 
a reduced description in which the electronic system (including
the couplings between subsystems) of interest
 are treated separately from the lattice vibrational 
modes at the site of the molecular  pigments.
The  Redfield procedure employing the   Born-Markov and secular 
approximations \cite{red,weiss,breu} and    the
non-Markovian approach  based on the 
 integro-differential equation  using  perturbation theory \cite{pot},
have been used to examine the quantum dynamical evolution of exciton states.
 A generalized Bloch-Redfield (GBR) equation technique \cite{caoIop} 
revealed  that temperature and spatial$-$temporal correlations in noise, partly arising from the vibrational motion of proteins, cannot be optimized simultaneously to yield the best
 energy propagation efficiencies. 
The collective system of exciton-vibronic modes 
was proposed as a likely reason for experimental observations of coherent oscillations \cite{Christ},
however this model was not supported by experimental work involving
doped samples of the FMO complex \cite{Hayes2011a}.

Based on the  quantum Markovian master equation of the Lindblad form, 
 the time evolution of the reduced open exciton state \cite{lind,gor} 
has been employed in some works \cite{caru09,caru10} with inclusion of
system-bath correlations \cite{johan,fas}. Noise which may arise due to static disorder,  dynamical stochastic fluctuations, or due to vibrational motion of proteins which are dependent on 
thermal effects in  non-equilibrium systems, is an  attribute which becomes favorable to the existence of quantum coherence. Optimal transport may occur at intermediate noise levels \cite{caru09,caru10}, 
with  very strong or weak noise levels resulting in lower
performances. Similar requirements are also essential for the stochastic resonance phenomenon to occur. These  conditions  also appear to be  consistent
with the level of noise perturbations required in the Grover-like search by excitonic states \cite{thilapra}.
In  earlier works on the influence of thermal noise \cite{caru09,caru10}, the disturbances were incorporated in  a 
local dephasing Lindblad master equation form within the site basis. 

 We note the  failings of the  Lindblad form \cite{sudar1,choi} for a range of temperatures and other dissipation factors.
To understand the weakness of the Lindblad form, we consider  
 a continuous semigroup  of linearly bounded operators,  ${\cal T}_t$, at time $t$,
which contracts  in the Hilbert space by virtue of the  trace and positivity preserving conditions.
Quantum states in general, can be associated with dynamical semigroups which form
the group of bounded operators satisfying: (i) ${\cal T}_0 = {\cal I}$, where {\cal I} is the identity,
(ii) ${\cal T}_{u+v}= {\cal T}_{u}+{\cal T}_{v}, \forall  u,v  \geq 0$, 
(iii) $||{\cal T}_t \rho || \leq ||\phi||$ for $t \geq 0$ for $\forall$ state $\rho$  in ${\cal H}$,
where $||A||=Tr[\sqrt{A^\dagger A}]$ and
 (iv) the existence of a continuous map $t  \rightarrow {\cal T}$ for  $\forall t > 0$. 
The set of  non-unitary contractive semigroup ${\cal T}_t$ in decaying states
is invariably linked to time irreversibility,  with 
 unitary (${\cal T}^\dagger = {\cal T}^{-1}$) and the completely non-unitary contractions may be 
analyzed using  the Langer-Sz-Nagy-Foias theorem (LSNF) \cite{langer}. The LSNF theorem states
that ${\cal T}$ can be decomposed into a unitary component ${\cal T}^u$ and
a non-unitary component ${\cal T}^n$, accordingly for every semigroup ${\cal T}_t$ ,
the Hilbert space ${\cal H}$ can be split  into two subspaces: ${\cal H} = {\cal H}^u \oplus {\cal H}^n $,
where ${\cal T}^u$ (${\cal T}^n$) is associated with ${\cal H}^u$ (${\cal H}^n$).

The similar focus of the  LSNF theorem and  Lindblad theory can be 
revealed by utilizing  the Hille-Yoshida generator for a dynamical semigroup
\cite{yosh,hille}. 
This generator yields a master equation (of the Lindblad form)
which  governs the evolution of $\rho$  \cite{lind,gor} :
\begin{eqnarray}
\frac{d}{dt}
\rho(t)= -{i}[{\cal H},\rho(t)]+\sum_{k,l=1}^{d}{\cal L}_{kl}(\rho(t))
\label{gks}\\
=-{i}[{\cal H},\rho(t)]+\frac{1}{2} \sum_{k,l=1}^{d}
a_{kl}\left(2 \chi_k\rho(t) \chi_l^\dag- \{\chi_k^\dag \chi_l,\,\rho(t)\}\right),
\label{seq}
\\  = - i [{\cal H},\rho(t)] +  \sum_{k=1}^{d} {\gamma_k}
\left( {\cal L}_k\rho(t) {\cal L}_k^\dag-\frac{1}{2} \{{\cal L}_k^\dag {\cal L}_k,\,\rho(t)\}\right)
\label{lind}
\end{eqnarray}
where the Lindblad operators, ${\cal L}_{kl}$ generate the map from the initial to the final density
operators of $\rho$ and ${\cal H}$ arises from a combination of the isolated system Hamiltonian,
${\cal H}_s$ and
a system-environment interaction operator. 
$\{\chi_k\}^{d}_{k=0}$ form the  basis in  the linear  operator space,  
with $\chi_0={\cal I}$, where ${\cal I}$ denotes the identity. The terms $(a_{kl})$  in Eq.~(\ref{seq})
 constitute the positive definite $d$-dimensional Hermitian 
Gorini-Kossakowski-Sudarshan matrix ${\cal A}$   \cite{gor}, with spectrum $\{\gamma_k\}$.
The first term in Eq.~(\ref{seq}) (or (\ref{lind})) represents reversibility in system dynamics,
and the symmetrized  Lindblad operators, ${\cal L}_k$ incorporate
environmental effects within the Born-Markov approximation and therefore
act as the source of  non-unitary dynamics.

The Lindblad form in Eq.~(\ref{lind}) ensures the positivity
of density operators at any time, however it is 
applicable only to weak to moderately weak system-reservoir couplings, when  Markov approximation
holds,  and breaks down in instances when  the
 complete positivity of density operators is  violated.
This could occur in the initial stage of evolution dynamics 
due to back flow of information from the reservoir bath at very short time
durations which are comparable to the bath memory times \cite{munro}. In these instances,
there appears to be a feedback action from the reservoir variables.
Time appears not to have a preferred direction in this regime, and  quantum  processes
therefore proceed (in the initial period) with some degree of  symmetry with respect to time reversal.
As a consequence, the finite time-scale of the vibrational environment  becomes relevant
during this quantum  dynamical regime. 

It may therefore be appropriate to use a non-Lindblad set of relations 
to describe the system dynamics  during the initial period of quantum evolution.
To this end, the non-perturbative hierarchical equations of motion (HEOM) technique
 \cite{hierac,tani,kreis} which  incorporates finite time scale of the dynamics in the vibrational 
environment, comes across as a viable tool in the study of ultrafast
quantum processes. The HEOM technique is based on a hierarchy of equations,  
in which the reduced density operator of the system couples to  a system of auxiliary operators.
These operators enable flexible tuning of  complex environments centered on 
 the correlation time of the bath, its spectral density and  spatial noise correlations. 
The HEOM model interpolates  between the Bloch-Redfield and the F{\"o}rster regimes,
 includes higher vibrational energies present in the 
environmental bath, and incorporates a second-order cumulant expansion that is exact for a harmonic
bath. 

A study   employing the HEOM technique \cite{kreis} showed that  the 
spectral density, which determines the decoherence and relaxation rate, played a critical
role in the  duration of coherent oscillations.
The results in Ref.\cite{kreis}  highlights  the links between strong coupling interactions 
required for fast thermalization processes and experimental observations 
of long-lasting coherent oscillations. In another work employing 
the HEOM technique \cite{kreis2},  the interplay between
electronic and vibronic degrees of freedom of the 
trimer model of the FMO complex was  used to highlight a likely
 underestimation of the life-time of electronic coherences captured by two-dimensional 
spectra results. More recently the HEOM technique \cite{guang} was used to investigate
 changes to transitions between sites and the protein environment, with results showing
robusticity of excitation propagation when defects are introduced into the molecular environment.
These findings are   of  particular relevance to the
 possibility that biological systems may adapt accordingly to achieve
 favorable characteristics to realize optimum outcome in 
its functionality. We will further examine the ability of biological systems
to adapt according to environmental changes using thermodynamic principles 
in Section \ref{thermo}.

The quasi-adiabatic path integral (QAPI) technique \cite{Makri,Thorwart}
 is another useful tool that provides reliable results at short reservoir correlation times, however it requires 
intense computational efforts at the low temperature regime. This approach is thus
unsuitable for structured reservoir systems that incorporate both narrow and broad spectral bands
In a recent work Kim et. al. \cite{kimkelly}, employed  an all-atom description 
of the photosynthetic complex within a semi-classical framework  to examine the 
Fenna-Matthews-Olson complex, and noted subtle differences in the role of
  vibrational modes at the ensemble and single-complex levels. Coherence was seen to be
weaker at the  ensemble level, and stronger at the single-complex level, while
thermal fluctuations in the chromophore couplings was seen to induce
 some level of redundancy in the coherent energy-transfer pathway \cite{kimkelly}.
 These results are interesting, as a semi-classical
framework was employed to show that the  coupling strength and  fluctuations within biomolecular systems
play important roles in the  coherent electronic transfer processes in photosynthetic systems.
We next describe the non-Hermitian quantum mechanical approach
from which other salient features, such as topological defects, can be revealed
during  energy propagation in organic systems.

\section{The non-Hermitian quantum mechanical approach}

It is well known that Non-Hermitian systems \cite{sense,thilherm,rotter,bend,fring,ali,bendJ,Heiss}
play important roles in the dynamics of open quantum systems. 
In particular, the appearance of non-Hermitian 
 terms (both real and imaginary)  have profound 
 implications for various physical and biochemical systems   modeled
as open quantum systems.
The striking difference between  non-Hermitian physics and  Hermitian physics lies in the
occurrence of degeneracies such as  exceptional points which are topological defects that occur when 
 two eigenvalues of an operator coalesce.
This may occur as a result of  changes in selected system parameters. Two mutually orthogonal  states
 then  merge into one self-orthogonal state,  resulting in a singularity in the spectrum \cite{Heiss}
with many intriguing effects. The critical parameter values  at which the singularity appear  are considered as
 exceptional points. These  points are known to be  located  in the vicinity
of a level repulsion \cite{Heiss} and unlike degenerate points,
only a single eigenfunction exists at the exceptional point.
There appears to be a decrease in information content due to the appearance
of the singularity in the spectrum, however the exact details of how information
is transferred to neighboring subsystems is currently not known.
  The  exceptional point remains to be observed  experimentally, however there has
been predictions that its existence during photosynthesis   may
  assist in distinguishing classical and quantum modes of transport \cite{thilherm}.

Quantum systems with non-Hermitian components  evolve  differently
 from  those of a   purely Hermitian Hamiltonian, with finite  lifetimes of states associated 
with attributes of 
the non-Hermitian Hamiltonian. The differences
between non-Hermitian and pure Hermitian nature of  Hamiltonians
can be examined using  the
 quantum brachistochrone concept, which is linked to 
 the minimum time taken to transverse the path  between two locations of
 a particle under a set of constraints. The passage time of  evolution of an
initial state  into the final state can be made arbitrarily small for
a time-evolution operator  which is  non-Hermitian but {\it PT}-symmetric \cite{bendJ}. This result
has  been generalized  to  non-{\it PT}-symmetric dissipative
systems \cite{fring}, with indications \cite{fring,bendJ}  that 
 propagation in non-Hermitian quantum mechanics  proceed  faster than those
of  Hermitian systems. The latter feature has  relevance to
problems examining the efficient propagation times noted in light harvesting systems.

The presence of  non-Hermitian attributes  also gives rise  to
 quasi-bound  resonance states   in the continuum reservoir partition \cite{rotter}.
While the  real and imaginary components  of the non-Hermitian 
 eigenfunctions evolve independently  during 
 avoided level crossing,  dynamical phase transitions arise 
due to the  distinct nature of the two components \cite{rotter}, along with
a bifurcation of the time scales  associated with the lifetimes of the resonance states.
Short-lived and  long-lived quantum states present during
non-Hermitian dynamics influence the dynamics of the
 non-ideal exciton  which result from  the Pauli exclusion principle
acting on  electrons and holes \cite{comb}
in material systems. Pauli exclusion based processes which operate
 independently of the well known Coulomb processes 
 in  fermion systems, contribute to hermiticity
of excitonic Hamiltonians \cite{thilma}. In a recent study \cite{thilma}, 
the non-ideal bosonic features of excitons was examined using
a non-Hermitian open quantum system approach.
Long-lived quantum coherence in photosynthetic
 complexes may be assisted by small bosonic deviation measures \cite{thilma} 
with involvement of a large number of excitons  during energy exchanges.
These conditions could help realize  a highly  correlated molecular environment conducive to 
efficient energy transport.

The rich dynamics inherent in non-Hermitian quantum  dynamics remains to be fully exploited,
especially in  studies involving the non-equilibrium quantum dynamics of dissipative molecular systems.
Yi et. al. \cite{yioh} recently introduced complex terms for the inter-site couplings for the 
Fenna-Matthews-Olson (FMO) complex, and obtained higher maximal energy transfer efficiencies
 compared to efficiencies linked to real inter-site couplings.
An earlier work \cite{thilherm} which utilized the
non-Hermitian two-level dimer model showed the persistence of
coherence features under specific environment conditions.
 Of  relevance to photosynthesis  is the 
demonstration that exceptional points may appear at critical temperatures for the 
dimer model \cite{thilherm}.
It appears that some
coherence features are retained during evolutions when
 the couplings of different subsystems to the environment becomes equal to each other \cite{thilherm}.
The subsystems of the dimer are seen to be unmeasured by the
environmental sources when there is 
equivalent couplings to the dissipation channels.
To this end, the indistinguishability of the sources of decoherence may result in the
preservation of   coherence during dynamics of the  photosynthetic dimer 
system.

\subsection{Quantum measurements during  photosynthesis}

 The subject of quantum measurements is a critical area of investigation  in the foundations of 
quantum mechanics, as it well known that   the interface  between theoretical models
and realistic outcomes (such as optical spectra) lies in direct measurements.
Quantum systems undergoing measurements have  counterintuitive properties,  with the state
of the system altered as a result of measurement.
The boundary between classical and quantum physics is  blurred
by quantum measurements.
In general, the results obtained in experiments are very much
 dependent on the instrument setup, and 
the quantity that is being determined and quantified via measurements. 
It is possible that other quantum signatures
may be present (aside from coherence oscillations) and has not, to date been detected
or observed due to the coarse resolutions of the monitoring instruments.

Quantum measurements   involve the use of  large-scale macroscopic devices to elucidate
 quantum features of a small-scale microscopic systems (e.g. photons). This procedure
is based on the assumption that superpositions present in a system of dimensions
 that scales several orders of 
magnitude smaller than that of the monitoring system, is transferred to the macroscopic measuring device.
It certainly appears rather challenging to  justify this assumption. Hobson   \cite{hobson} proposed
that in the case of an apparatus performing  ideal measurement that is able to 
distinguish between the distinct entities of a  superposed system, the composite
system-apparatus state becomes entangled by virtue of the Nonlocality attribute.
Thus the unitary evolution of the global composite state is left intact, and 
there occurs  coherent transfer of the original 
superposition present in the observed system to the global superposed state.
These results are verifiable using nonlocal two-photon interferometry experiments  \cite{hobson},
and can be used to improve tomographic measurements of fragile quantum
 systems \cite{tomog,tomoprl,excitomo} such as natural light harvesting molecular
networks.

The deep links between quantum measurement and the peculiar features of
biosystems that yield enhanced energy transfer
efficiencies is a promising  area for future studies, in view
of advancements in tomography based  reconstructions of the
density matrix of entangled systems \cite{tomoprl}.
Through tomography measurement techniques \cite{tomopres}, it may be possible
to map out the global 
 density matrix of multipartite states  in large biomolecular
systems. The inclusion of quantum measurement principles is expected to assist in the
 examination of correlation measures, particularly in the vicinity of the
exceptional point,   and in the understanding
 of the quantum dynamics of light harvesting systems.

\section{Non-Markovianity  and photosynthesis}{\label{mar}}

Non-Markovianity is a property that is intricately linked to the violation
of the trace preserving, completely positive (CP) mapping attributes of 
quantum dynamical semigroups \cite{sudar1,choi}. The mathematical maps in Markovian environments,
 possess divisibility and  provide tractability when 
characterizing the evolution dynamics of memoryless quantum systems. As mentioned in Section \ref{other},
this numerical tractability was the key reason for its use to improve models employed 
in earlier works \cite{red} which examined
the dynamics of a quantum system in contact with  external reservoir.
 Non-Markovian dynamics dominates 
 in the presence of strong system-environment coupling regime or when un-factorized initial
conditions exist between the system and environment. 
 Pechukas \cite{per} demonstrated  that  non-Markovianity 
may be present as an artifact of the product of the initial conditions, 
 $\rho_s(0) \otimes \rho_r(0)$, 
where $\rho_s$ ($\rho_r$) denote the density operator specific to the system, $s$ (reservoir, $r$).
There seems to be a  non-observance of a statistical interpretation for the reduced dynamics of
the quantum system as a result of non-Markovian dynamics. While it is known that non-Markovianity 
arises as a result of the interconnectedness of  past and future events and  
 quantum interferences,  the origins of this peculiar attribute in quantum
systems still  remains unresolved.  

Current measures of non-Markovianity are generally based on the deviations of  existing quantum structure 
from the continuous, completely positive semi-group attributes of Markovian evolution
characterized by the  dynamical map $\rho(0)\mapsto\rho(t)=\Phi(t,0)\rho(0)$.
Due to varying characteristics of different distance
 measures (e.g. trace distance, Bures distance, Hilbert-Schmidt distance \cite{niel}), 
there is no unique  quantifiable measure of non-Markovianity in quantum systems. 
The distinguishability attribute quantified by the 
decrease in the trace-distance,   $D[\rho_1,\rho_2]$ between two system states,
 $\rho_{1},\ \rho_2$ is well known measure,  that does not increase under all completely positive, 
trace preserving maps \cite{breu2009}.
Accordingly,  $\sigma=dD[\rho_1,\rho_2,t]/dt$  assumes  respective negative or positive values,
 when information flows outwards or into  a system,  that is coupled to  its environment.
The increase of trace distance during 
any time interval is therefore taken as a signature  of  non-Markovianity \cite{breu2009}.
An alternative  measure is based on divisibility \cite{wolf,Rivas2010}, which
incorporates the characteristics of quantum correlations of the ancilla component
 of  an entangled system evolving under  a trace preserving completely
 positive quantum channel.  There has been several
 discussions related  to the equivalence of these two 
measures of non-Markovianity, with the  overall agreement that the underlying generic features
of the quantum system  remains a critical factor for the two measures to be reconciled.
There are other measures, some of which that are easily computable, introduced in recent years,
which we omit discussing in greater details here. 
Nevertheless, the myriad of ways that non-Markovianity can be defined and examined
only serves to highlight the challenges in adopting a  rigorous study
of the contributions of non-Markovian dynamics to coherent oscillations in photosynthetic systems.

We consider briefly well known studies on the role of non-Markovianity in enhancing the photosynthetic activities
of biomolecular systems. Several works have shown that
 the environment noise (both Markovian and non-Markovian) can enhance the propagation of  energy  in light 
harvesting systems ~\cite{lloyd,caru09,caru10,thilzeno},
with the fine interplay between quantum coherence and environmental noise 
determining the optimal  functionality of photosynthetic  systems.
Non-Markovian processes are generally prominent in the  reorganization energy regime \cite{caoIop}
where the energy transfer efficiency is optimized. This may be due to 
preservation of coherences along critical pathways, as 
non-Markovian effects have been noted to
increase the lifetime of entanglement effects \cite{caru09,caru10}.
The exciton entanglement dynamics of the 
 Fenna-Matthews-Olson (FMO) pigment-protein  complex in an earlier work \cite{thilchem2} also showed 
 increased oscillations of entanglement  in the non-Markovian
regime. Violations of forward time translations may
interfere with the  optimal balance of quantum and incoherent dynamics
required for efficient energy transfer, however this possibility needs further
 verification via rigorous numerical computations using realistic parameters of photosynthetic systems.

\begin{figure}
\centering
\includegraphics[width=9cm]{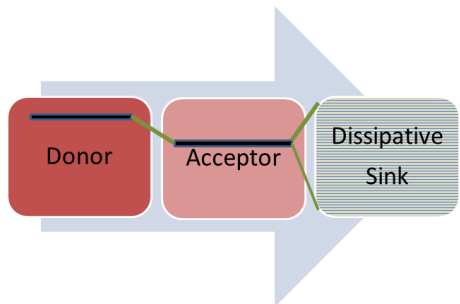}
\caption{Simplified model  of the donor and acceptor protein pigment complexes with
 discrete energy levels, in which the acceptor is linked to a dissipative sink reservoir
with continuous frequency spectrum. Energy from the acceptor flows into the reservoir with time.
}
 \label{sink}
\end{figure}
Recently  we examined the non-Markovian dynamics in a
 multipartite system of two initially correlated atomic qubits \cite{thilusha}, 
with each qubit placed in a single-mode leaky cavity coupled to a  bosonic reservoir. 
The atom-cavity-reservoir system  is analogous to the photosynthetic
model constituting the donor and acceptor protein pigment
complexes, with the latter coupled to a  third entity represented by the  phonon dissipative
sink (see Fig. \ref{sink}).  This work \cite{thilusha} showed 
the dominance of non-Markovian features in several two-qubit partitions,
with maximal  non-Markovianity  present in the cavity-cavity
subsystem  in the vicinity of the exceptional point. 
It was noted that the fidelity of the cavity-cavity partition
experiences a minimum at this unique topological  point \cite{thilusha}.
On the basis of these results,  segments of biomolecular chains  appear to
act as  quantum channels, with the local minima in the fidelity measures 
possibly arising from the  merger of the two eigenvalues at the
topological defect. Further investigations are needed
to confirm the roles of molecular chains as part of an intricate
quantum network model system endowed with some degree of quantum information processing
abilities, and to examine the unique roles of exceptional points in these systems.

In another related study on photosynthetic systems \cite{thilzeno},
 the time domains involved during effective 
Zeno or anti-Zeno dynamics appeared to be of the same order of 
magnitude as the non-Markovian time scale  of the reservoir correlation dynamics.  
An earlier study \cite{thilzeno} highlighted
the subtle links in decay rates due to the Zeno mechanism  and 
information flow between specific   partitions of entangled systems. These
findings have  implications for a joint Zeno effect-non-Markovian
action in  the critical tripartite states of the 
donor-acceptor-sink model of photosynthetic biosystems.

\section{Coherence and Entanglement}

Despite the wide range of techniques employed in  modeling studies and experimental
observations of oscillatory structures,  processes
of classical origins and those with an intrinsically quantum mechanical
coherence origins have not been fully demarcated in light harvesting systems.
The detection of  the transition point at which a classical world departs to give way
to one dominated by quantum effects remains unresolved, even though 
there are   distinct differences between 
classical and quantum systems. There exist no
transformation in which a   quantum world  can be mapped onto a  larger region of deterministic outcomes,
however it can be conjectured that elements which are uniquely quantum 
vanish beyond a critical system size.  Experimental results point to the
possibility that  photosynthetic systems possess a greater content of 
quantum correlations, and for this reason, the role of coherence and entanglement
entities play a critical role in their dynamics in noisy conditions.

The appearance  of quantum entanglement allows
the maximal knowledge of a composite system, but does not come with the freedom to  assign a 
specific state to subsystem without discarding its links to other subsystems.
This includes one of several  puzzling properties of 
quantum states, with  quantum entanglement  noted
as a valuable resource for the implementation of quantum computation
and quantum communication protocols \cite{niel,horo,horo96}. These protocols  include 
quantum teleportation \cite{ben93,tele2,pop}, 
dense coding \cite{wies}, quantum cryptography \cite{hill99,fuch} and
 remote state preparation \cite{remote}. Entanglement as an entity is capable of
continual destruction and regeneration in a composite system-environment setup \cite{woot},
even if no prior correlation existed between the subsystems. In some
systems, entanglement may vanish at a certain  finite time  prior to the decay 
of the coherences \cite{death1,death2,death3,dur,zhong}. These issues
are a topic of interest, as 
entanglement as a resource is meaningless if it vanishes rapidly.

\begin{figure}
\centering
\includegraphics[width=9cm]{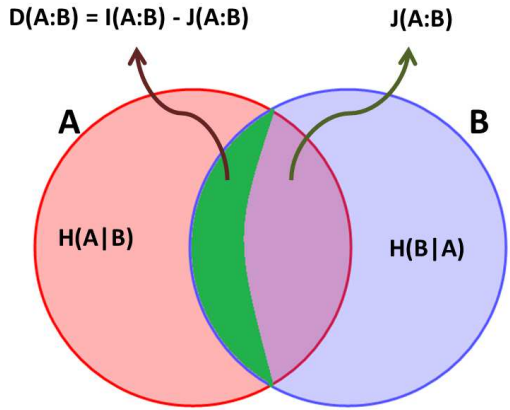}
\caption{Relationship between conditional quantum entropies $H(A|B)$ and $H(B|A)$,
mutual information $I(A:B)$,
the locally accessible classical correlation $J(A:B)$ and the 
inaccessible information quantified by the quantum discord $D(A:B)$.
$J(A:B)$ is the maximized content obtained by measuring $B$ \cite{zu,ve1,ve2}.
}
 \label{discord}
\end{figure}
Separable or non-entangled states may possess
 other kinds  of non classical correlations  such as 
 the quantum discord \cite{zu,ve1,ve2}, 
based on the difference between quantum and classical information content in
entangled systems (see Fig. \ref{discord}). The quantum  discord is more robust than entanglement, and  is not
susceptible to sudden death occurrences. The discord entity
can also be created by local operations on the   measured system
via non-unital channels \cite{cicca12}. Interestingly,  in the case
of  two-qubit states undergoing non-dissipative decoherences, it is possible
that the discord remains
resilient under certain conditions during non-Markovian evolutions \cite{froz}.
The occurrence of a constant quantum discord has been demonstrated in an
earlier work on the quantum processing attributes of J-aggregate systems \cite{thilJ}.
There are other interesting properties of this measure, such as, two positive discord states
can be mixed to obtain a zero-discord classical state, and
zero-discord classical states in  orthogonal directions
can be merged to form a non-zero discord state \cite{matt}.
The quantum discord is not restricted by the monogamy rule \cite{woot}
which is a requirement for the concurrence measure during  entanglement sharing.

The  intriguing features of quantum discord  described in the earlier paragraphs,
 have potential  applications  within the non-Markovian environment of
light harvesting systems. The discord measure as well as other non classical correlations may 
be used to examine the   occurrences of quantum phase transitions and 
 quantum  communication protocols in biomolecular systems. 
It has been noted that for all known quantum information protocols dependent on quantifiable
two-qubit discord,   there exist noisy evolutions in the state space for which coherence
is preserved and where communication protocols are unaffected by noise \cite{lombo}.
The theoretical framework under which these result were obtained \cite{lombo}, may be easily extended
to examine  the presence of strong coherences,
in  the noisy environments  of photosynthetic systems. 

There have been few genuine attempts to link entanglement and the  quantum discord
 measure to processes which sustain bio-cellular activities at
 physiological temperatures \cite{pdavis}. There  
are obvious difficulties to these efforts, as the
relationship between coherence and entanglement 
remains  subtle,  with no  known demonstrable explicit link between the two entities. 
Decoherence  in open quantum
systems may take an infinite amount of time to vanish unlike entanglement.
This may be due to factors such as the degree of robustness of the initial state
 of the quantum system under study, and presence of non-Markovian dynamical entities.
It is highly likely that
robustly entangled states as initial states are more resistant to decoherence processes
than uncorrelated states at the start. Thus the number of molecular sites excited by
photons, as well as the intensity of illumination at the antenna sites, are key factors
involved in further understanding the coherence phenomena in biosystems.

\section{Multipartite states in light harvesting systems}
Entanglement or non-classical correlations that spans over several lattice sites,
may be represented by multipartite states. For instance, a 
qubit state $\Psi$ associated with $N > 2$ subsystems  is present in the multipartite state as
\be
|\Psi\> = \sum_{n=1}^N c_{n} \left( |0\>^{\otimes (n-1)}\otimes|1\>\otimes|0\>^{\otimes (N-n)} \right)
\ee
 with  coefficients  $c_n$, and $|0 \rangle, |1\rangle$ are orthonormal basis vectors of a two-dimensional state space.
The  symmetric  Dicke state with just one excitation appear as
\be 
\label{wn}
|W_N \> =\frac{1}{\sqrt{N}} \left(|100...0\>+|01...0\>+...+|0...01\> \right),
\ee
while the GHZ states  are obtained as
\bea
\left\vert GHZ_{N}\right\rangle &=&\alpha \underbrace{ | 0 \rangle \otimes | 0 \rangle \otimes \cdots | 0 \rangle }_{N}
+\beta \underbrace{ | 1 \rangle \otimes | 1 \rangle \otimes \cdots | 1 \rangle }_{N} \\ \nonumber
&=&
\alpha\left\vert 0\right\rangle ^{\otimes N}+\beta\left\vert 1\right\rangle ^{\otimes N}
\eea
where  $\left\vert \alpha\right\vert
^{2}+\left\vert \beta\right\vert ^{2}=1$. 
Multipartite states   possess a  rich source 
of local and nonlocal correlations due to the
multitude of partitions available within a group of entangled qubits, and 
have relevance  in the realistic situation  of an entire photosynthetic membrane constituting 
many FMO complexes and thousands of bacteriochorophylls. Such a structural platform
facilitates the formation of a  large cluster of massively entangled excitonic qubits.
Multipartite states  are generally fragile compared to bipartite states, and in this regard,
appear unlikely to play a dominant role in the presence of 
decoherence effects in organic systems. However there have been predictions that   such states
can exist even at physiologically  high
temperatures \cite{temp,thilchem2} in solids,  and   play an important
role in the quantum properties of molecular systems.

Multipartite states are also of interest in  photosynthetic systems,
from the point of the ``principle of quantum information causality", founded on the mathematical
formulations of  quantum information causality \cite{pital}. This principle states that 
the maximum amount of quantum information that can be transferred from one state to another,
is bounded by the quantum system's dimension, and is not reliant
on any physical resources previously shared by the communicating states. 
The dimension of a quantum system here, refers to the number of different possible pathways
available to a measured  system. This means that light harvesting systems with
large dimensions hold higher quantum  communication capacities, and almost
certainly,  multipartite states are favored participants of quantum information processing
during photosynthesis.

Unlike bipartite states,  the examination of  the robustness of quantum states 
is a  challenging task in multipartite states, as the entanglement and non-classical
correlations  are not to easy  categorize for such states. There exist no simple route
 to  the specification of entanglement or quantum discord in
multi-state quantum systems. There are variations in the robustness of different types 
of multipartite states (Dicke, GHZ, W or cluster states) under decoherence processes
as noted in earlier studies \cite{gohn,borras}. 
For instance, the W state  is known to be 
highly robust as a   multipartite  entangled state, with respect to loss 
of a single excitation \cite{cirac}. However  it is uncertain
 whether the  W state is more  robust against decoherence
than the GHZ states, mainly due to the incompatibility of comparison of the dynamics of the two types
of states during a decoherence process. The W states belong, in the case of  the  three qubit
 symmetric pure states, to the group characterized by two distinct Majorana spinors \cite{ushraj}.
On the other hand, the GHZ states belong to the group characterized by three distinct Majorana spinors.
In the case of  GHZ states which undergo decoherence \cite{aol}, the 
entanglement decays 
 faster when there is increase in the number of initially entangled particles.
Differences between the  GHZ and  W states also exist
in terms of their polygamous nature, for instance,  the generalized W states can be mono or polygamous
while the generalized GHZ states exhibit only monogamy features with respect to the quantum deficit measure \cite{ushraj}.

We  have examined the exciton entanglement dynamics of the 
 Fenna-Matthews-Olson (FMO) pigment-protein  complex from the green sulfur bacteria
of the species, \textit{Prosthecochloris (P.) aestuarii} \cite{lorenExpt} using typical
 values of the reservoir characteristics at cryogenic and physiological temperatures \cite{thilchem2}. 
The light-harvesting system was considered as a 
global system constituting several smaller subsystems  interconnected via  quantum correlations, where
the  important contribution of specific tasks such as  
teleportation \cite{ben93,tele2} and   quantum state splitting \cite{split1,split2} 
within  a noisy environment was demonstrated. 
Quantum communication protocols
such as  quantum teleportation \cite{ben93,tele2}, 
and quantum secret sharing \cite{split2}, 
are utilized by multipartite states   to assist energy transfer during 
photosynthetic processes \cite{thilchem2}. In
largely extended light harvesting  systems with
intricate network connectivity, multipartite states appear robust
with respect to decoherence processes \cite{thilchem2}. In particular,
 quantum information processing
involving teleportation followed by the decodification
tasks in W states of the FMO complex may  account for
experimental results which show persistence of coherent oscillations
at physiological temperatures.

Recently, a protocol known as quantum energy teleportation (QET) \cite{hotta}
 was proposed to show the viability of energy teleportation between two remote sites.
Through this scheme, a quasi-particle 
is able to effectively transfer energy to another point in space via local operations 
and classical communication. 
The elegance of this proposal  \cite{hotta}
 lies in the fact that it is information instead 
of tangible energy that  is being transferred from one point to another.
As a consequence there is no violation 
in energy conservation nor generation 
of heat during the propagation of
quantum information. This principle may be extended to examine teleportation processes
 in photosynthetic systems,
where information transfer between specific sites results in extraction of energy from the surrounding phonon bath
environment. The interplay between information and energy in correlated biomolecular systems
has implications for the  possible role of quantum thermodynamical principles, which we 
examine in the next Section.

\section{Quantum thermodynamics,  energy and information}\label{thermo}

Thermodynamic principles are traditionally used to track with accuracy,
processes that involve  exchanges of  energy and work in classical systems.
The translation of similar schemes to quantum systems has been an intense area of investigation 
in the last few years \cite{hov,kaw,kaw2,Saga1,Saga2,op,arne,gros}, 
despite the well known
links between information theory and thermodynamics, based on 
the parallels between  Shannon's uncertainty
function and the entropy function. 
This is due to the challenges involved in categorizing work in terms of
 entanglement, mutual information, quantum discord and other probabilistic
measures of correlations (both classical and quantum). As a consequence of the varying characteristics of 
different correlation measures, there appears to be 
 no single quantifiable measure, or a unique definition of work or energy dissipation in
quantum systems.  The identifiable measures of work in  quantum systems  have a 
statistical distribution linked to various evolution routes taken by  the quantum systems. In this regard
the work or energy transformations that occur in quantum systems stand distinct from analogous
measures which appear in classical systems. Interestingly, the
work measure in quantum systems has a rich structure attached to it by the 
inherent link between a probabilistic interpretation of work or energy, and 
 quantum statistical mechanics in the non-equilibrium regime. These reasons form the key motivation
for considering  quantum thermodynamic factors in light harvesting systems,
which can be viewed as  operating in  the ``far-from-equilibrium" regime.

Initial studies of  links between thermodynamics and quantum correlations
commenced  with demonstration of the association
 between energy and information theory, 
 in the context of computation energy cost \cite{laud}.
Landauer's erasure principle showed that any irreversible  process expands  work due to the entropy transfer 
from the degrees of freedom,  coded as information entities outwards into the environment \cite{laud}.
Decades earlier, Szilard had used the  Maxwell's demon model to show that $k_B T\ln 2$ of work 
can be extracted from a thermodynamic cycle, and highlighted that a positive entropy 
production in measurement compensates for the work gained during the cycle \cite{Szilard}. This ensures that 
the second law of thermodynamics is left intact. This idea has recently been used
to compute the binding energies of composite boson systems \cite{thilszil},
and  may be extended to examine non-ideal photosynthetic excitons
from a quantum informative-theoretic perspective.

The extension of thermodynamic principles to non-equilibrium conditions is a very challenging task.
The mathematics of non-equilibrium statistical dynamics appear tractable only in the near-equilibrium 
regime with approximately linear structures. 
To this end, the entropy production fluctuating theorem \cite{crooks}
differs from several mathematical formalisms in its applicability to states that are
operating in the far from equilibrium regime. This theorem stems from
the Jarzynsky relation \cite{jarzy} used to
examine  states which undergo dynamical evolutions in the unstable regimes.
The Jarzynski relation \cite{jarzy} yields a neat and tractable
relationship between the distribution of work
performed on a classical system by an external force and
the free energy difference between the initial and final
states. The significance of the Jarzynski relation
lies in the fact that quantities at equilibrium, such as  energies
between equilibrium states, are linked to  non-equilibrium paths associated with 
 measurements.  Though the relation has been obtained for classical systems,
it holds equally well for quantum systems, as demonstrated  recently \cite{shaun2}.

The principles of the Szilard engine has been  extended  via the fluctuation theorem \cite{crooks} to
the formulation of thermodynamic work based  on a time forward and backward shifting technique \cite{kaw}.
The average dissipated work $\langle W\rangle_{\rm diss}$ required to translate  a 
system from one canonical equilibrium state to another one without change in 
temperature $T$ is obtained as \cite{kaw,Jar}
\begin{equation}
\label{wk}
\langle W\rangle_{\rm diss}=\langle W\rangle-\Delta F= k T \left \langle \ln
\frac{\rho}{\tilde{\rho}} \right\rangle.
\end{equation}
$\langle W\rangle_{\rm diss}$ is therefore the additional work, other than
the difference in free energy $\Delta F$, 
that is required for
the transition process. $\rho=\rho(\Gamma;t)$ is the
 probability  density associated with observing the system  in the
micro-state $\Gamma=(q,p)$, at time $t$
 for position (momentum) $q$ ($p$).
$ \left \langle \; \right\rangle$ denotes the averaging of $\rho$,
and  $\tilde{\rho}=\tilde{\rho}(\tilde{\Gamma};t)$ is the time-reversed
distribution  observed 
at phase point $\tilde{\Gamma}=(q,-p)$  for the same
duration as the forward process. 

The dissipated work~(\ref{wk}) is also obtainable 
in terms the relative entropy, $D(\rho\|\tilde{\rho})$ between $\rho$ and $\tilde\rho$  \cite{kaw2}
\begin{equation}
\label{main2}
\langle W\rangle_{\rm diss}= k T D(\rho\|\tilde{\rho}),
\end{equation}
where
 $D(\rho\|\tilde{\rho})$ 
is the relative entropy between $\rho$ and $\tilde\rho$. From Eq.\ref{main2},
the gain in total entropy (system plus heat
bath) in the forward process, $\Delta S$, is given
 by \cite{kaw2}
\begin{equation}\label{entr}
\Delta S=k  \left \langle \ln
\frac{\rho}{\tilde{\rho}} \right\rangle = k D(\rho\|\tilde{\rho}).
\end{equation}
The classical result in Eq. \ref{entr} can be seen in  quantum settings
via extension of  $D(\rho\|\tilde{\rho})$ to its quantum counterpart
based on the relative entropy \cite{ved,shaun2}.

 Unlike the classical
entropy measure, the quantum entropy obeys 
the sub-additivity property, which results in subsystems
acquiring a greater entropy  than  the whole system.
Eq. (\ref{main2}) reveals the deep links between
dissipation and the non-Markovianity measure. As pointed out in Section \ref{mar}, the latter
is linked to non-divisible maps in systems which undergo non-contractive quantum evolution and
can be represented by the  relative 
entropy, $D(\rho\|\tilde{\rho})$. We note that $D$  is essentially  a
distance measure. This  means that the quantity of dissipated work,
$\langle W\rangle_{\rm diss}$ {\it decreases} 
with time, during  a positive Markovian process.
Conversely, the rate of increase in the dissipated work at $t' > t$
 is  sufficient but not a necessary signature of non-Markovianity. Other works \cite{Saga1,Saga2} have 
shown that the work  extracted from a system is determined by the  mutual information present in 
the  system-environment configuration, and quantum correlations can be quantified based 
on thermodynamic principles \cite{op}. 
In a study on multipartite systems \cite{hov}, the connection between 
entanglement generation and work power was shown,  with optimal work extraction  
achievable without generation of any form of entanglement. However it is possible
that entanglement can be created with extraction of a large
 power output \cite{hov}. It would be worthwhile to pursue these ideas
within the correlated environment of molecular systems in future investigations
involving quantum thermodynamics aspects of photosynthesis.

\subsection{Predictive Power}

Recently, a vivid and rather profound view of the system-environment dynamics was given in terms
of the system's ability to predict its future interactions
with one or more sections of its surroundings in classical systems  \cite{Still}. 
In the case of living systems, the   ability of the 
 biological system to ``foretell" future events  becomes fundamental to increasing the survival of 
the supported species, as it also enhances the success rate of various tasks needed for cellular functioning.
The change in system state as part of its response 
to a changing environment may be  interpreted as a form of implicit computation that is performed by the system.
This gives rise to un-factorized states of existence of the system-environment system
due to temporal correlations at all times (past and future).
In systems that possess predictive power  \cite{Still},
 it was shown that the information theoretic measure of the inefficiency of the
 predictive process (the non-predictive information) can be equated to thermodynamic inefficiency. 
This inefficiency is based on the  work dissipated during the duration that the system  evolves from
one point to another. The ideas proposed by Still et. al. \cite{Still} highlight that
the effective use of information and
efficient thermodynamic operation are intertwined operations, and best studied
as  dual measures during the evolution of a system.

The connections between predictive power of information entities and  thermodynamic processes
has been  extended to quantum systems in a very recent work  \cite{arne}. It was
shown that the change in system entropy, conditional on the presence of  relevant environmental variables, 
is a measure of the computation's thermodynamic inefficiency based on  the entropy change \cite{arne}
\bea
 \beta W_{\rm lost}[\rho_{SX}\to\rho_{SX}'] &=& \ln 2[H(S|X') - H(S|X)] \\
&= & \ln 2 [I(S:X) - I(S:X')],
\eea
where $\beta = 1/k_B T$ denotes the inverse temperature, while the meanings of the entropy terms ($H(S|X'),H(S|X),I(S:X)$)
are made explicit in Fig. \ref{discord}. $S$  denotes the finite
quantum system, while  $X$ is a subset of the surrounding environment that is correlated with $S$.  $X$ evolves via
 \be
\rho_X \to \rho_X' = \mathcal{E}(\rho_X)
\ee
where $\mathcal{E}$ denotes a quantum channel.

Grimsmo \cite{arne} employed a simple quantum model 
to illustrate that the entropy change is a measure of ineffective use of information \cite{arne},  
consistent with the idea demonstrated earlier for  classical systems \cite{Still}. 
Environmental coherence of the central system was shown to improve predictability to an amount
 quantified by a negative (quantum) constituent of the dissipated work \cite{arne}.
The thermodynamic inefficiency of the most energetically efficient classical approximation of a quantum  system
provides an operational interpretation to  quantum discord.
Thus for the case of two correlated systems, $A$ and $B$, the non-predictive quantum information 
held by $B$ equals its lost work potential if the system $A$ changes state, and in the 
case where  $A$ assumes a classical state, the lost work is given by the quantum discord present
 in $B$ prior to the changes in A. Therefore the quantum discord can be seen to quantify 
the reduction in work potential under an optimal classical approximation of some part 
of the system surroundings \cite{arne}.

In an earlier work \cite{gros},  the operational definitions of the quantum and classical
 correlations in a bipartite quantum state was given in terms the amount of work needed to erase various
types of correlations. In particular, the work required to erase the quantum correlation is
one that will culminate in the appearance of a separable state.
The mutual information was specified in terms of the minimal work
needed to reduce a bipartite system to a product state, whereby all traces of correlations are eliminated  \cite{gros}.
These ideas may be extended to examine  work that is dissipated in select regions of
biomolecular systems due to dynamical stochastic fluctuations occurring in other regions of the 
 photosynthetic membrane  holding the  bacteriochorophylls. With a similar focus, changes
in work or energy in select segments of biomolecules due to enviromental fluctuations
arising from the  vibrational motion of protein molecules in spatially
separate regions may also be assessed. Future investigations may consider the detailed examination
of  optimized conditions involving temperatures, solar illumation levels
 or structural arrangements of the molecular sub-units,
that yield minimal  dissipation of energy placed at the initial site of excitation.

Systems with high predictive powers have implications for  the semigroup law and 
  contractions in the Hilbert space of 
Hilbert-Schmidt operators,  with likely  violations of forward time translations. 
Non-Markovianity and quantum predictability may  have common origins, and the investigations of possible common
grounds will contribute to further understanding of the  thermodynamic operations of biomolecular 
systems during photosynthesis. We emphasize that the  concept of a system's predictive power with respect
to a changing environment, remains to be tested within a non-Hermitian framework.
In this regard, there are ample opportunities to seek out new findings within the
myriad of models where non-Hermitianity can manifest itself, aside from
biomolecular systems.

\subsection{ Far from equilibrium regime in light harvesting systems}

Based on the discussions in the earlier sections, it is clear that thermodynamic principles
from an  information theoretic perspective, has  practical relevance in the highly correlated environment
 of light harvesting  systems. As the photosynthetic process involves a sudden creation 
of an initial excitation, the biomolecular system
is transferred to a ``far from equilibrium" operating regime. The interconnected biomolecules
and the surrounding  environment undergo rather chaotic fluctuations determined by external parameters (temperature, solar illumination).
The  non-equilibrium dynamics involves a complex interplay of system-environment dynamics and 
 non-Markovian predictive-type back-action of the  environment on the system.
Depending on the spectral profile of the environmental bath, there is rapid shift to an equilibrium
state (short bath memory) or a non-equilibrium state  of system-environment (long bath memory).

We note that  the  Crooks \cite{crooks} and  Jarzynski equality relations \cite{jarzy} are valid for open quantum
systems \cite{shaun2}, and are independent of the system-environment coupling strength
and discrete structures present in the 
 thermal environment. The quantum versions of the fluctuation relations therefore
appear  suitable to be employed in investigating the quantum environment of biomolecules
at higher temperatures. The latter  systems which  undergo
``far-from-equilibrium" fluctuations,  may incorporate useful information based on predictive powers,
that is distinct from the background noise. 
The system's predictive ability \cite{Still}
may have enabled some robustness against decoherence,  reflected by
the comparatively long duration of strong coherent oscillations observed in spectroscopy experiments. 
It is expected that parameters linked to any form of structure within the
 environments other than white noise, may  be critical 
to  the embedded system's predictability, with influence on the 
 complicated interplay between different entities of 
quantum correlations. The embedded system itself
may use its predictive abilities to  tune or even select favorable characteristics
 of the environment giving rise to fluctuations that  result in optimized transport  properties.
 Quantum systems may ``select"  a range of parameters when there is 
match in time scales of the system and its environment.
Such control action gives rise to a pattern of evolution  based on  a   continuous process 
of realizing optimality in critical activities within a biological network of biomolecules 
embedded in chaotic environments.
This may occur during the ``far from equilibrium" dynamical evolution  of photosynthetic
systems.

The thermodynamic efficiency appear as a viable measure of the ability of molecular systems 
to dissipate  minimal solar energy, while transporting
the reserve energy from the receiving end to the destination site
where energy conversion  takes place.
The operational efficiency highlights the need to comprehend
 the  fundamental quantum thermodynamics processes at work, and closely examine the 
factors that govern the energy conversion at the reaction center \cite{mclare}.
Light harvesting systems viewed
as quantum thermodynamic optimal machines also provide a reliable
framework from which  salient observations can be made,
and applied to other structures which are independent specific biochemical features
such as neural network structures.
Schneider \cite{schne} has highlighted the use of Shannon's theorem for communication channels 
to describe ``machine capacity" of molecular systems. The  availability of the complex code of a
 molecular machine's operation improves its efficiency (by reducing
the frequency of malfunctioning). While dissipation processes and thermal noise
restricts the ``machine capacity", there is opportunity to increase the  chances of 
survival   by careful control of the accuracy of
information exchanges between the distinct components constituting the organism.
These ideas are expected to provide further insight  into one of nature's best nano-machines,
and reveal vital clues to fabrication of artificial quantum molecular  systems.  

\subsection{Agent-based modeling and simulation approach of photosynthetic biomolecular systems}

A scenario where  biomolecules ``think" and ``act" like intelligent
 biological networks \cite{schne,hess,whit,gross} may
appear far fetched, however the presence of these attributes may be investigated using an
agent-based modeling and simulation approach \cite{agent0,agent1,agent2}. 
Tasks that require some self-learning
may be formulated as the optimization of 
expected outcomes (such as the high conversion efficiencies)
within the framework of quantum correlations that operate in some
specific  subspaces,  otherwise considered as domains of  agents.
In the simplified model of the green sulfur bacteria,
each FMO monomer may be viewed as an independent unit endowed with quantum correlations that 
mimic the role of interacting agents. These agents may obey a set of basic rules as well as 
self-learning rules that adapt to a changing environment, and incorporate
 quantum memory which may coded as classical or non-classical correlations. 
Overall, the  agent seeks to achieve a set of goals, and in the case of photosynthetic biological
structures, this would be to maximize energy transfer with the optimum efficiency
that is achievable at the prevailing environmental conditions.
While the  agent is considered to act  independently, the further extensions
of these ideas to an emergent behavior attributed to a collection
of multi-agents may be examined in simulation studies of  light harvesting systems.
Furthermore there is motivation to pursue the links
between agent-based rules and the quantum predictive power proposed
in Ref. \cite{arne}, to examine the optimality of biomolecular 
structural layouts for energy transport and other information processing \cite{schnei} properties.

\section{Conclusions}
In the preceding sections, we have discussed various topics of importance
to studies of  quantum correlated dynamics in natural light harvesting systems.
The last ten  years has seen a surge of interest in
this topic, with agreements and conflicting results between different research
groups. This should not be  surprising as there are differing interpretations, and
 approaches of undertaking   quantum mechanical studies and extracting
quantum structures from  highly complex systems of
molecular structures. There have been numerous studies with various perspectives and 
terminologies that has been proposed in the literature, and we remind readers
that only some references have been represented in this review. And the
challenges of dealing with some seemingly basic foundations of quantum physics,
must be held in view, when quantum information-theoretic
interpretations are made of experimental
results of light harvesting systems.

Two critical  measures 
based on the efficiency  of energy propagation, and robustness in quantum coherence
appear vital to photosynthetic processes. While the energy transfer efficiency is linked
to the overall network connectivity and structural configuration of biomolecules,
the strength of quantum coherence  appears to be dependent on
the system's response to a changing environment. There is a fine interplay 
between two seemingly counteracting  effects, one based
on maintenance of structure while the second is dependent on a changing environment.
This review has examined some possible ways to 
 unify these core functions of the  light harvesting process, by employing the elements of
quantum informative protocols and quantum thermodynamical principles. The examination of the links between  the predictive power,
 non-Markovianity  and thermodynamic quantities presents  a 
rigorous and  unified approach that could  provide deep insights to understanding
nature's mode of energy propagation. Future investigations using these
ideas will help seek  comprehension of the links between quantum rules and
nature's design of biological network structures, which constitute
living forms which are able to harvest solar energy with remarkable efficiencies.

\section{Acknowledgments}

 The author gratefully acknowledges the  support of  the Julian Schwinger Foundation Grant,
JSF-12-06-0000, and access to National Computational Infrastructure
(NCI) facilities at the Australian National University.  The author would like to thank Monique Combescot, Malte Tichy and
Alex Bouvrie for 
 correspondences on  composite  boson systems, and Susanne Still and  A. L. Grimsmo
for insightful discussions  related to predictive information and
 thermodynamic inefficiencies  of quantum systems.


\end{document}